\documentclass[letter,bibyear]{aa} 

\usepackage{epsfig}
\usepackage{amsmath}
\usepackage{subfigure}
\usepackage{natbib}
\usepackage{multirow}
\usepackage{color}
\usepackage{longtable}
\usepackage{graphicx}
\usepackage{epstopdf}
\usepackage{booktabs}
\usepackage{txfonts}

\newcommand{\cpl}{Chem. Phys. Lett.}
\newcommand{\chp}{Chem. Phys.}
\newcommand{\jms}{J.~Mol.~Spectrosc.}   
\newcommand{\jmst}{J.~Mol.~Struct.}   

\newcommand{\cd}{cm$^{-2}$}
\newcommand{\kms}{km s$^{-1}$}
\newcommand{\once}{10$^{11}$\,cm$^{-2}$}
\newcommand{\doce}{10$^{12}$\,cm$^{-2}$}
\newcommand{\trece}{10$^{13}$\,cm$^{-2}$}
\begin{document}

\title{Discovery of CH$_2$CHCCH and detection of HCCN, HC$_4$N, CH$_3$CH$_2$CN, and, tentatively, CH$_3$CH$_2$CCH in TMC-1
\thanks{Based on observations carried out with the Yebes 40m telescope
  (projects 19A003, 19A010, 20A014, 20D023).
  The 40m radio telescope at Yebes Observatory is operated by the Spanish Geographic Institute
  (IGN, Ministerio de Transportes, Movilidad y Agenda Urbana).}}

\author{
J.~Cernicharo\inst{1},
M.~Ag\'undez\inst{1},
C.~Cabezas\inst{1},
N.~Marcelino\inst{1},
B.~Tercero\inst{2,3},
J.~R.~Pardo\inst{1}, 
J.~D.~Gallego\inst{2},
F.~Tercero\inst{2},
J.~A.~L\'opez-P\'erez\inst{2}, and
P.~de~Vicente\inst{2}
}

\institute{Grupo de Astrof\'isica Molecular, Instituto de F\'isica Fundamental (IFF-CSIC), 
C/ Serrano 121, 28006 Madrid, Spain. \email jose.cernicharo@csic.es
\and Centro de Desarrollos Tecnol\'ogicos, Observatorio de Yebes (IGN), 19141 Yebes, Guadalajara, Spain.
\and Observatorio Astron\'omico Nacional (IGN), C/ Alfonso XII, 3, 28014, Madrid, Spain.
}

\date{Received; accepted}

\abstract{
We present the discovery in TMC-1 of vinyl acetylene, CH$_2$CHCCH,
and the detection, for the first time in a cold dark cloud, of HCCN, HC$_4$N, and CH$_3$CH$_2$CN.
A tentative detection of CH$_3$CH$_2$CCH is also reported.
The column density of vinyl acetylene is (1.2$\pm$0.2)$\times$10$^{13}$ cm$^{-2}$, which makes it
one of the most abundant closed-shell hydrocarbons detected in TMC-1. Its abundance is only three times lower than that of propylene, CH$_3$CHCH$_2$. 
The column densities derived for HCCN and HC$_4$N are
(4.4$\pm$0.4)$\times$10$^{11}$ cm$^{-2}$ and (3.7$\pm$0.4)$\times$10$^{11}$ cm$^{-2}$, respectively.
Hence, the HCCN/HC$_4$N abundance ratio is 1.2$\pm$0.3. For ethyl cyanide we derive a column density 
of (1.1$\pm$0.3)$\times$10$^{11}$ cm$^{-2}$. 
These results are compared with a state-of-the-art chemical model of TMC-1, which is able to account for the 
observed abundances of these molecules through gas-phase chemical routes.
}


\titlerunning{CH2CHCCH and exotic cyanides in TMC-1}
\authorrunning{Cernicharo et al.}

\maketitle

\section{Introduction}
The chemical complexity of the interstellar medium has
been demonstrated by the detection of more than 200 different chemical
species. In this context, the cold dark core TMC-1 presents an interesting carbon-rich chemistry that led to
the formation of long neutral carbon-chain radicals and their anions
(see \citealt{Cernicharo2020a}, and references therein). Cyanopolyynes, which are stable molecules,
are also particularly abundant in TMC-1 (see \citealt{Cernicharo2020b,Cernicharo2020c}, and references therein).
The chemistry of this peculiar object produces a large abundance of
the nearly saturated species CH$_3$CHCH$_2$, which could mostly be a typical molecule of hot cores
  \citep{Marcelino2007}. 
  The polar benzenic ring C$_6$H$_5$CN has also been detected for the first time
    in space in this object \citep{McGuire2018}, while benzene itself has only so far been 
been detected toward post-asymptotic giant branch objects \citep{Cernicharo2001}.

Sensitive line surveys are the best tools for unveiling the molecular
content of astronomical sources and searching for new
molecules. 
A key element for carrying out a detailed analysis
of line surveys is the availability of exhaustive spectroscopic information
of the already-known species, their isotopologues, and their
vibrationally excited states.
The ability to identify the maximum
possible number of spectral features leaves the cleanest
possible forest of unidentified ones, therefore opening up a chance
to discover new molecules and get insights into the chemistry
and chemical evolution of the observed objects.
In this context we have recently succeeded in discovering,
prior to any spectroscopic information from the laboratory, 
several new molecular species (see, e.g., \citealt{Cernicharo2020a,Cernicharo2020b,Cernicharo2020c,
Cernicharo2021a,Cernicharo2021b}; \citealt{Marcelino2020}).
Consequently, the astronomical object under study has become a real spectroscopic laboratory.
Experience shows that, if the sensitivity of a line survey is sufficiently high,  many unknown species
will be discovered, providing key information on the ongoing chemical processes in the cloud.

In this letter we report on the discovery of vinyl acetylene, CH$_2$CHCCH, a species that has been 
spectroscopically characterized
but never detected in space. We also present the detection, for the first time in a cold dark cloud, of 
HCCN, HC$_4$N, and CH$_3$CH$_2$CN. Ethyl acetylene, CH$_3$CH$_2$CCH, has also been tentatively detected.

\section{Observations}
\label{observations}
New receivers, built within the Nanocosmos project\footnote{\texttt{https://nanocosmos.iff.csic.es/}}
and installed at the Yebes 40m radio telescope, were used
for the observations of TMC-1. The Q-band receiver consists of two high electron mobility transistor cold 
amplifiers covering the
31.0-50.3 GHz range with horizontal and vertical polarizations. Receiver temperatures vary from 17 K at 32 GHz
to 25 K at 50 GHz. 
Eight 2.5 GHz wide fast Fourier transform spectrometers, with a spectral resolution
of 38.15 kHz, provide the whole coverage of the Q-band in each polarization.
The main beam efficiency varies from 0.6 at
32 GHz to 0.47 at 50 GHz. A detailed description of the system is given in \citet{Tercero2021}.

The line survey of TMC-1 ($\alpha_{J2000}=4^{\rm h} 41^{\rm  m} 41.9^{\rm s}$ and $\delta_{J2000}=+25^\circ 41' 27.0''$)
in the Q-band was performed over several sessions. 
Previous results obtained from the line survey 
were based on two observing runs, one performed in November 2019 and one in February 2020.
They concerned the detection of C$_3$N$^-$ and C$_5$N$^-$
\citep{Cernicharo2020b}, HC$_5$NH$^+$ \citep{Marcelino2020}, HC$_4$NC \citep{Cernicharo2020c}, and HC$_3$O$^+$
\citep{Cernicharo2020a}.
Additional data were taken in October 2020, December 2020, and
January 2021 to improve the line survey and to
further check the consistency of all observed spectral features. These new data allowed the detection
of HC$_3$S$^+$ \citep{Cernicharo2021a} along with the acetyl cation, CH$_3$CO$^+$ \citep{Cernicharo2021b}, the isomers of C$_4$H$_3$N \citep{Marcelino2021}, and HDCCN \citep{Cabezas2021}.

Two different frequency coverages were used in the line survey, 31.08-49.52 GHz and 31.98-50.42 GHz, to ensure that no spurious spectral ghosts were produced 
in the down-conversion chain, 
which down-converts the signal from the receiver to 1-19.5 GHz and then splits it into eight 2.5 GHz bands, which are
finally analyzed by the FFTs. 
The observing procedure was frequency switching with a frequency throw of 10\,MHz for the first two runs and
8\,MHz for the later ones. The intensity scale (i.e., the antenna temperature, $T_A^*$) was calibrated 
using two absorbers
at different temperatures and the atmospheric transmission model  \citep[ATM;][]{Cernicharo1985, Pardo2001}. Calibration
uncertainties were adopted to be 10~\%. The nominal spectral resolution of 38.15 kHz was kept for the final
spectra. The sensitivity varied across the Q-band from 0.5 to 2.0 mK. All data were analyzed using the GILDAS
package\footnote{\texttt{http://www.iram.fr/IRAMFR/GILDAS}}.

\section{Results and discussion} 
\label{results}
The sensitivity of our observations toward TMC-1 (see Sect.~\ref{observations}) is a factor of 10-20 better than
previously published line surveys of this source at the same frequencies \citep{Kaifu2004}. This large
  improvement allowed us to detect a forest of weak lines, most of which arise from the isotopologues
of abundant species such as HC$_3$N, HC$_5$N, and HC$_7$N \citep{Cernicharo2020c}. 
In fact, it was possible to detect many individual lines \citep{Marcelino2021} from molecular species
that had previously only been reported using stacking techniques.
Taking into account, on one hand, the large abundances found
in TMC-1 for cyanide derivatives of abundant species and, on the other, the presence of nearly saturated hydrocarbons 
such as CH$_3$CHCH$_2$ 
\citep{Marcelino2007}, we searched for species such as CH$_2$CHCCH and CH$_3$CH$_2$CCH,
as well as for cyanides found previously only 
in carbon-rich stars (HCCN and HC$_4$N) or in warm molecular clouds (CH$_3$CH$_2$CN).
Line identifications in this TMC-1 survey were performed using the MADEX catalogue \citep{Cernicharo2012}, 
the Cologne Database of Molecular Spectroscopy (CDMS) catalogue (\citealt{Muller2005}), and the JPL. 

\begin{figure}[]
\centering
\includegraphics[scale=0.6,angle=0]{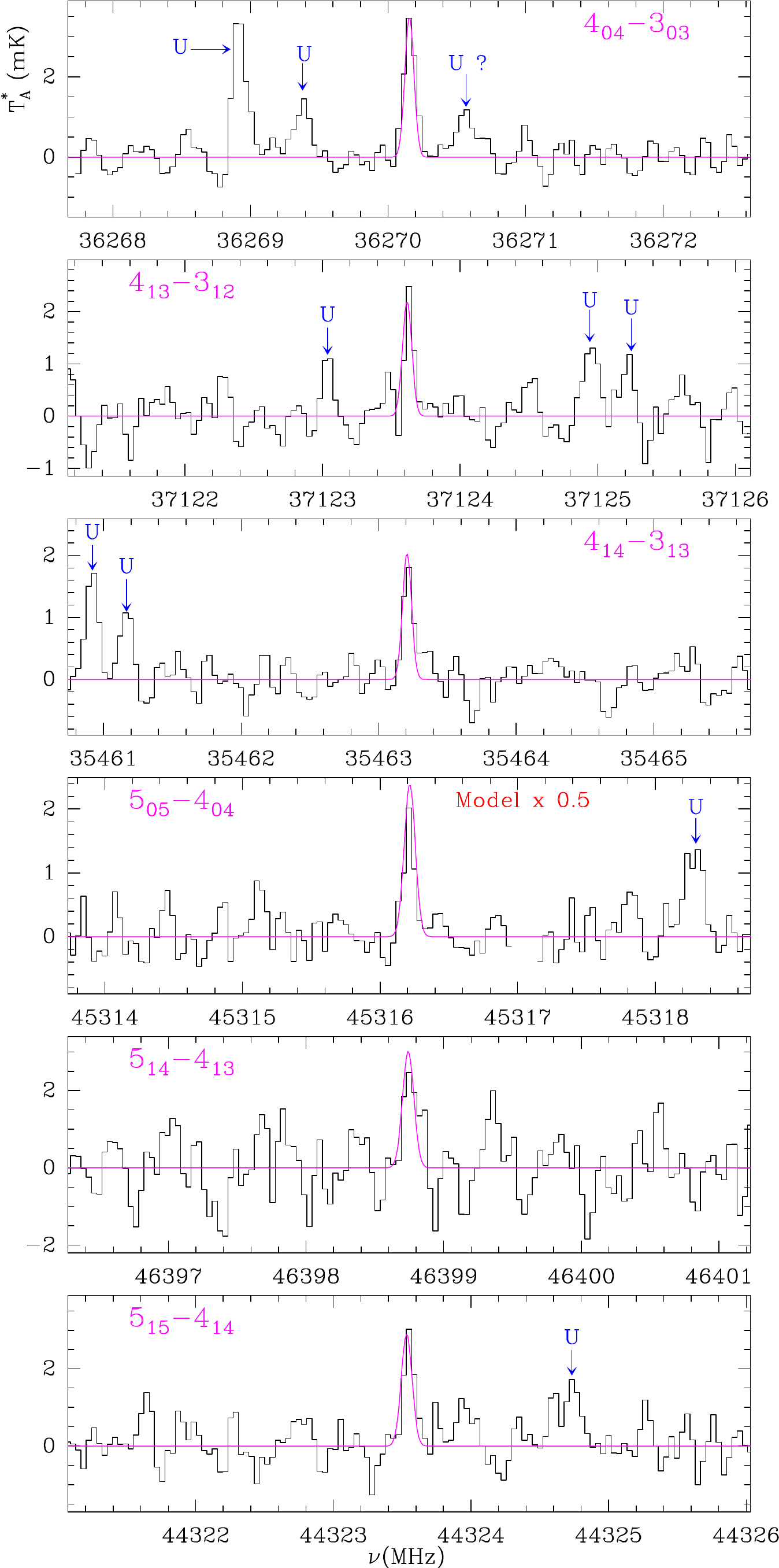}
\caption{Observed transitions of CH$_2$CHCCH toward TMC-1.
The abscissa corresponds to the rest frequency of the lines assuming a v$_{LSR}$
for the source of 5.83 km s$^{-1}$. Frequencies and intensities for the observed lines 
are given in Table \ref{tab_line_parameters}. 
The ordinate is the antenna temperature, corrected for atmospheric and telescope losses, in millikelvins (mK).
The purple line shows the synthetic
spectrum obtained for this species.
}
\label{fig_ch2chcch}
\end{figure}

\begin{figure}[]
\centering
\includegraphics[scale=0.58,angle=0]{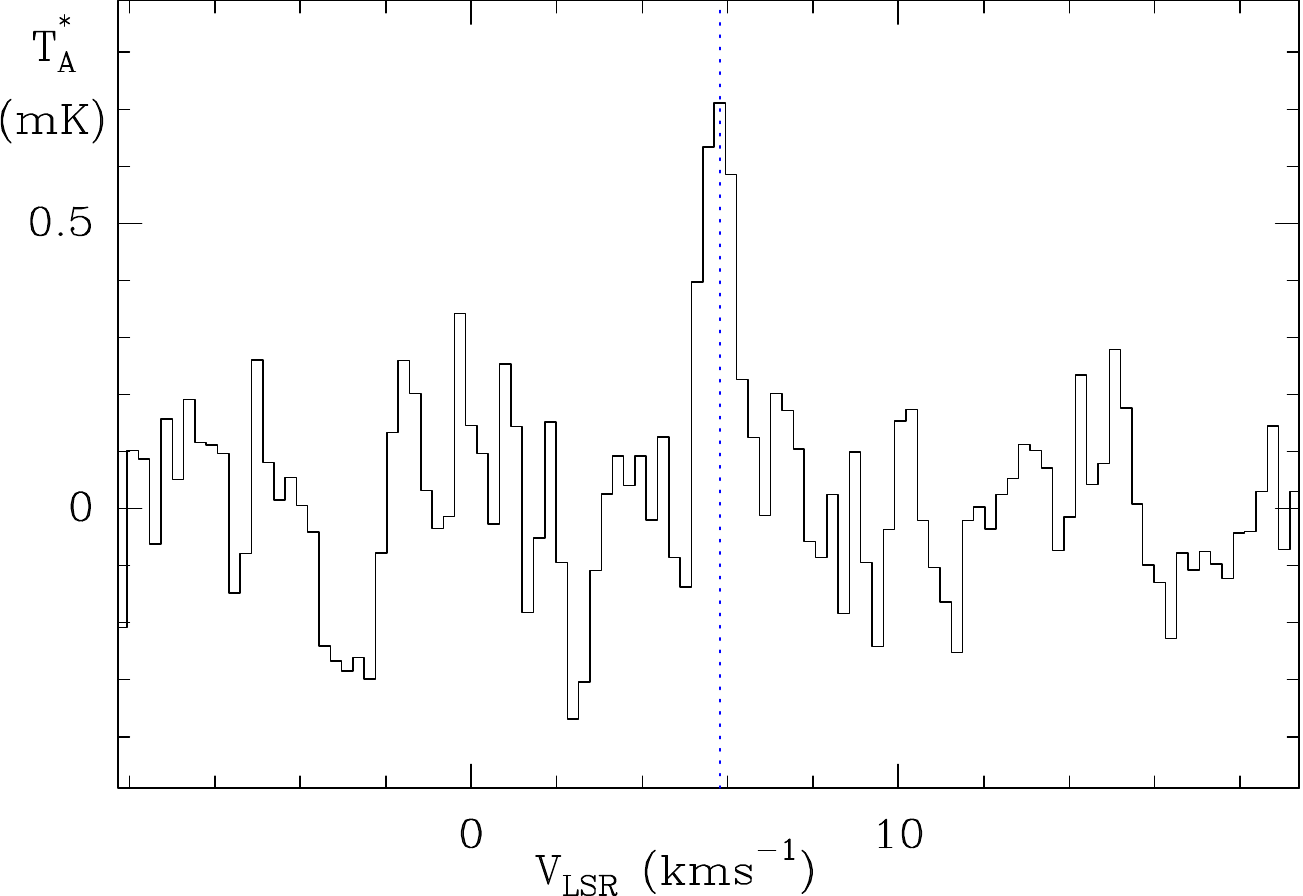}
\caption{Line profile for CH$_3$CH$_2$CCH toward TMC-1 from
  stacking the signal from the transitions of
  this species in the 31-50 GHz frequency range.
The abscissa corresponds to the local standard of rest velocity. 
The ordinate is the antenna temperature, corrected for atmospheric and telescope losses, in mK.
The observed spectra from the individual lines are shown in Fig. \ref{fig_ch3ch2cch}.
}
\label{fig_ch3ch2cch_sta}
\end{figure}

\subsection{Vinyl acetylene, CH$_2$CHCCH}
\label{CH2CHCCH}
Spectroscopic constants for CH$_2$CHCCH were derived from a fit to the lines reported by \cite{Thorwirth2004} and
implemented in the MADEX code \citep{Cernicharo2012}. We detected six lines with $K_a$=0 and 1 (see Fig.~\ref{fig_ch2chcch}).
Although the observed line intensities (see Table \ref{tab_line_parameters}) are
low ($\sim$2-3 mK), the column density for this molecule should be particularly large as its dipole moment, $\mu_a$,
is only 0.43 D \citep{Sobolev1962,Thorwirth2003}. Only a $K_a$=2 line was marginally detected, corresponding to the
$4_{22}-3_{21}$ rotational transition.  
Assuming a uniform source with a radius of 40$''$ \citep{Fosse2001} and through a standard rotational diagram,
we derived a rotational temperature of 5.0$\pm$0.5 K
and a column density for vinyl acetylene of (1.2$\pm$0.2)$\times$10$^{13}$ cm$^{-2}$. 
Figure \ref{fig_ch2chcch} shows the synthetic spectrum of CH$_2$CHCCH
computed for the derived rotational temperature and column density (see Appendix \ref{line_parameters}). The 
comparison with the data shows a very good
agreement for all lines, with the exception of the 5$_{05}$-4$_{04}$ transition, for which the predicted intensity is nearly a 
factor of two larger than what is observed. This line is probably affected by a negative spectral feature resulting from the
folding of the frequency switching data. 

Using the column density of
H$_2$ derived by \citet{Cernicharo1987}, the abundance of CH$_2$CHCCH relative
to H$_2$ toward TMC-1 is 1.2$\times$10$^{-9}$. This abundance is only three times lower than that of propylene \citep{Marcelino2007}, 
and ten times lower than that of methyl acetylene \citep{Cabezas2021}.
Hence, vinyl acetylene is one of the most
abundant hydrocarbons in TMC-1 and probably the most abundant compound containing four carbon atoms. It is interesting
to compare the abundance of vinyl acetylene with that of vinyl cyanide. We analyzed the lines of CH$_2$CHCN
covered by our Q-band data. We derived a rotational temperature of 4.5$\pm$0.3 K,
and $N$(CH$_2$CHCN)=(6.5$\pm$0.5)$\times$10$^{12}$ cm$^{-2}$
(see Appendix \ref{CH2CHCN}). Hence, the abundance ratio between the acetylenic and cyanide derivatives of
ethylene (CH$_2$CH$_2$) is 1.8$\pm$0.4, which suggests that ethylene could be a likely precursor of CH$_2$CHCCH and CH$_2$CHCN through 
reactions with CCH and CN, respectively (see Sect.~\ref{discussion} for more details).

\begin{figure}[]
\centering
\includegraphics[scale=0.58,angle=0]{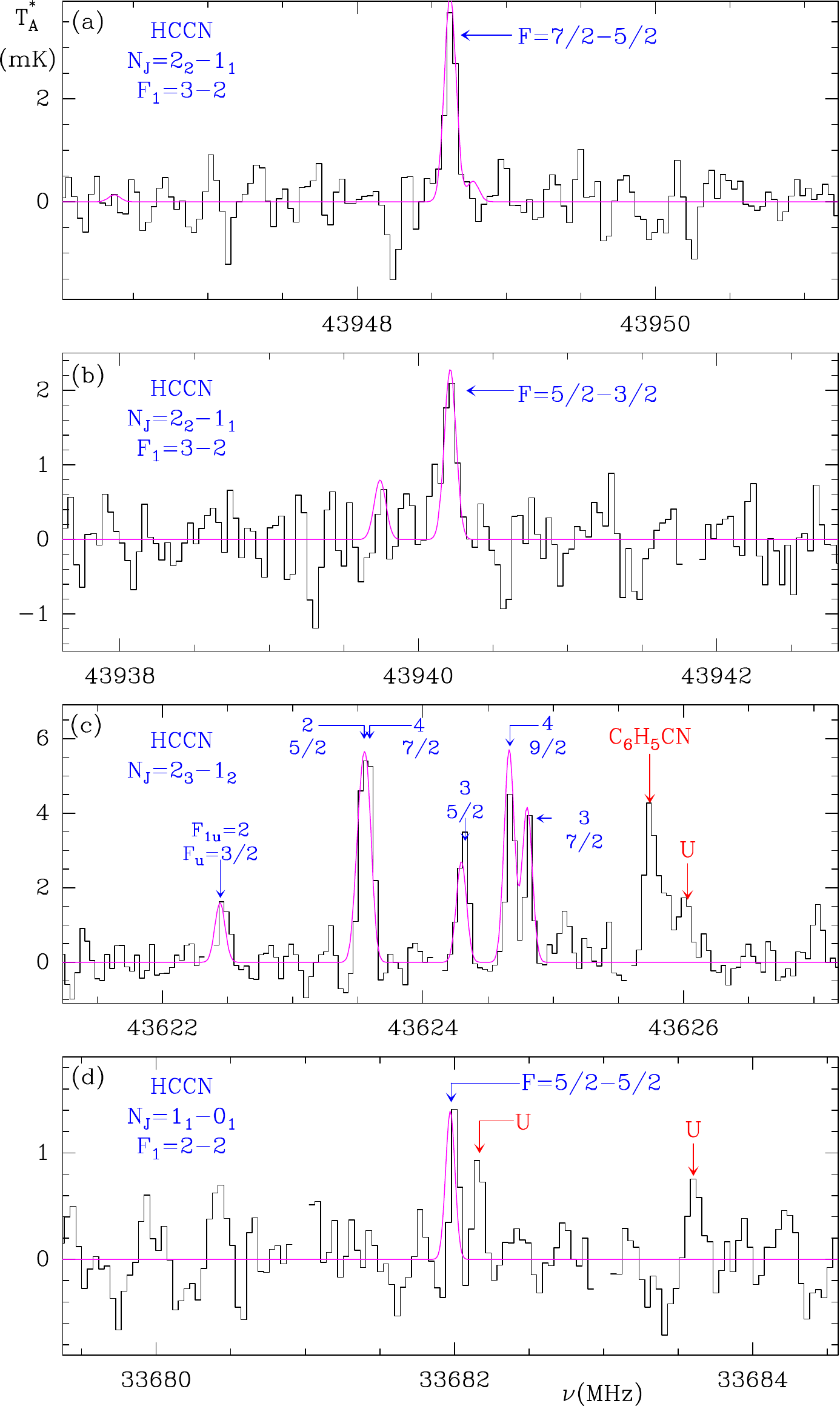}
\caption{Observed lines of HCCN in our Q-band survey toward TMC-1.
  The abscissa corresponds to the rest frequency assuming a v$_{LSR}$ of
  the source of 5.83 km s$^{-1}$.
The ordinate is the antenna temperature, corrected for atmospheric and telescope losses, in mK. The violet
line shows the synthetic spectrum computed for T$_r$=4.5 K and N(HCCN)=4.4$\times$10$^{11}$ cm$^{-2}$.
For the sake of clarity, only the upper quantum numbers F$_1$ and F are provided in panel (c).
}
\label{fig_hccn}
\end{figure}

We searched for CH$_3$CH$_2$CCH in our survey. Figure \ref{fig_ch3ch2cch_sta} shows the resulting spectrum
from stacking the lines of this species covered in our data (see details in Appendix \ref{CH3CH2CCH}). 
The whole set of individual lines is shown in Fig. \ref{fig_ch3ch2cch}. We consider this molecule to be tentatively detected with a column density of 
9$\times$\once. 
We also detected ethyl cyanide, CH$_3$CH$_2$CN, in TMC-1 (see Sect. \ref{CH3CH2CN}) with a column density of \once. 
Hence, assuming that CH$_3$CH$_2$CCH in TMC-1 is detected, the abundance ratio CH$_3$CH$_2$CCH/CH$_3$CH$_2$CN is $\sim$9. 

\subsection{HCCN and HC$_4$N}
\label{HCCN}
The cyano methylene radical, HCCN, was detected toward the carbon-rich star IRC+10216 by \citet{Guelin1991}.
The molecule has been observed in
the laboratory by several authors \citep{Saito1984,Endo1993,McCarthy1995,Allen2001}. Its dipole moment is 3.04 D
\citep{Inostroza2012}. Given the large abundance of cyanopolyynes in TMC-1 (e.g., \citealt{Cernicharo2020c}), 
HCCN could be expected in this source, although it was searched for by \citet{McGonagle1996} with no success. 
Several of its rotational transitions, which show significant hyperfine splitting, are within 
the frequency range of our TMC-1 survey. All of them were detected, as shown in Fig. \ref{fig_hccn}
(see Table \ref{tab_line_parameters}). 
This is the first time HCCN has been detected in cold interstellar clouds. 
The range of level energies covered by the observed transitions, 1.6-3.7 K, did not allow us to
build a reliable rotational diagram. We adopted the rotational temperature derived for CH$_2$CHCN 
(4.5$\pm$0.5 K; see Sect.~\ref{CH2CHCCH}) and computed a synthetic spectrum with the column density as a free parameter. 
The best fit was obtained for a column density of (4.4$\pm$0.4)$\times$10$^{11}$ cm$^{-2}$. As shown in Fig. \ref{fig_hccn},
the match between the model and the observations is very good.

\begin{figure}[]
\centering
\includegraphics[scale=0.54,angle=0]{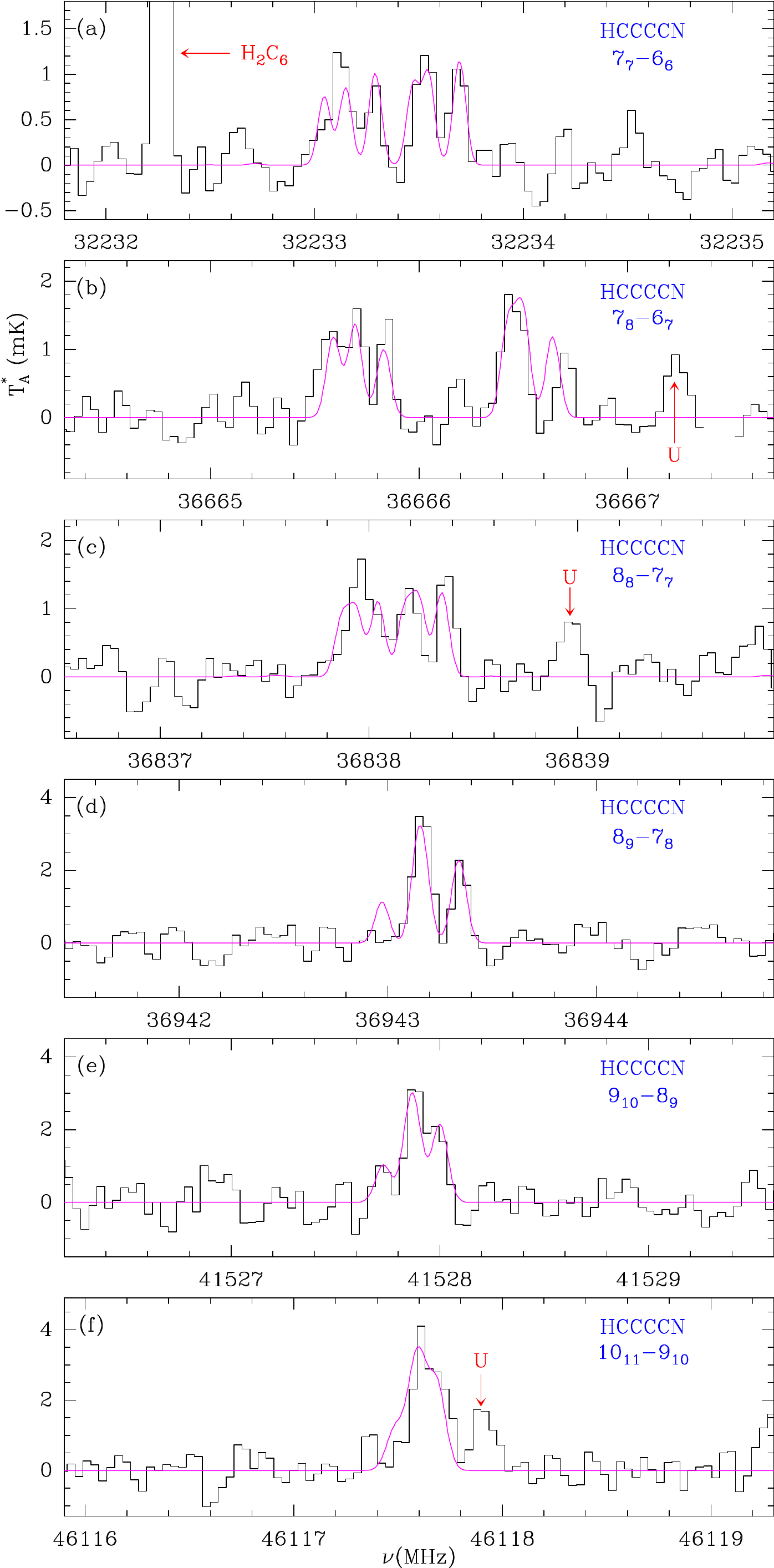}
\caption{Selected lines of HC$_4$N in the 31-50 GHz frequency range toward TMC-1.
The abscissa corresponds to the rest frequency assuming a v$_{LSR}$ for TMC-1 of 5.83
km s$^{-1}$. 
The ordinate is the antenna temperature, corrected for atmospheric and telescope losses, in mK.
The violet line shows the synthetic spectrum computed for \mbox{T$_r$=4.0\,K} and N(HC$_4$N)=3.7$\times$\once. 
}
\label{fig_hc4n}
\end{figure}

In addition to HCCN, the linear cyano ethynyl-methylene radical, HC$_4$N, was also detected in IRC+10216 
\citep{Cernicharo2004}.
Spectroscopic information on this molecule \citep{Tang1999} was implemented in MADEX. Its dipole moment is 4.33 D
\citep{Ikuta2000}. The observed lines toward TMC-1 are shown in Fig. \ref{fig_hc4n}.
From those lines and the model fitting procedure, it is possible to derive a rotational temperature
of 4$\pm$1 K and a column density of (3.7$\pm$0.4)$\times$10$^{11}$ cm$^{-2}$. Figure \ref{fig_hc4n} shows the
synthetic spectrum corresponding to these parameters. The agreement with the observations is
very good. This is the first time that this species has been detected in a cold dark cloud. 

We searched for 
the bent and cyclic isomers of HC$_4$N \citep{McCarthy1999a,McCarthy1999b} without success.
We derived 3$\sigma$ upper limits to their
column densities of 1.5$\times$\once\, and 1.0$\times$\once, respectively. 
It is interesting to note that these two isomers have a very different molecular 
structure relative to linear HC$_4$N \citep{Cernicharo2004}. Hence, they are probably not formed in the 
reaction between CH and HC$_3$N, which is the main pathway to the linear isomer.
Finally, we searched for the related radicals CCN \citep{Kakimoto1982,Ohshima1995} 
and C$_4$N \citep{McCarthy2003}. However, due to their low permanent dipole moment
\citep{Pd2001,Fiser2013}, we obtained
very conservative 3$\sigma$ upper limits to their abundances of 1.8$\times$\doce\, and 4.0$\times$\trece\,, respectively.

The abundance ratio HCCN/HC$_4$N derived in TMC-1 is 1.2$\pm$0.3, which is very different than that 
derived in the carbon-star envelope IRC\,+10216 ($\sim$ 9; \citealt{Cernicharo2004}). In IRC\,+10216, this abundance ratio is a 
factor of two larger than the HC$_{2n+1}$N/HC$_{2n+3}$N decrement observed for cyanopolyynes 
\citep{Cernicharo2004}, while in TMC-1 the HCCN/HC$_4$N ratio is two to three times lower than that of the cyanopolyyne decrement. The 
different behavior 
in TMC-1 compared to IRC\,+10216 is probably caused by differences in the 
abundances of the precursors of HCCN and HC$_4$N (see Sect.~\ref{discussion}).

\subsection{CH$_3$CH$_2$CN}
\label{CH3CH2CN}
This molecule is typical of warm molecular clouds, where it produces a forest of lines arising from all its
isotopologues and low-energy vibrational excited states \citep{Demyk2007,Daly2013}.
It was searched for in TMC-1 by \citet{Minh1991} without success. More recently, \citet{Lee2021} used
stacking techniques, providing an upper limit to its column density in TMC-1 of 4$\times$\once.
We searched for the lines of this molecule in our TMC-1 Q-band survey. Figure \ref{fig_ch3ch2cn} shows the six lines
with $K_a$=0 and 1 detected. All of them appear at 4-5$\sigma$ levels. The $K_a$=2 lines are too weak
to be detected with the sensitivity of our survey.

We computed a synthetic spectrum
using the rotational temperature derived for CH$_2$CHCN (4.5 K; see Appendix \ref{CH2CHCN})
and a column density of $N$(CH$_3$CH$_2$CN)=1.1$\times$\once\, (with an estimated error of 30\%). The match between the observations and 
the model is very reasonable.
We derived abundance ratios CH$_2$CHCN/CH$_3$CH$_2$CN=65$\pm$20 and CH$_2$CHCCH/CH$_3$CH$_2$CCH=120$\pm$40.
It is interesting to compare the abundance ratio between vinyl cyanide and ethyl cyanide in TMC-1 and that in Orion KL.
In the latter source 
it is 0.06 \citep{Lopez2014}, which is a factor of $\sim$1100 smaller than in TMC-1. 
The spatial distribution of both molecules in Orion KL has been analyzed by \cite{Cernicharo2016}. Both species arise
in regions with temperatures of 100 and 350 K.
The huge difference in this abundance ratio tells us
about the different chemical processes prevailing in cold and warm molecular clouds. While the
chemistry is dominated by the contribution of the evaporating ices covering dust grains in objects such as Orion KL, reactions between radicals and neutrals in cold dark clouds
play a key role in producing these molecules.

\subsection{Chemistry of detected molecules}
\label{discussion}
After the discovery of abundant propylene in TMC-1 \citep{Marcelino2007}, the detection of an 
even larger, partially saturated, and abundant hydrocarbon, such as vinyl acetylene, in the same 
cloud brings to light the existence of a rich organic chemistry in cold dark clouds, going 
beyond the long-known presence of unsaturated carbon chains. Moreover, just as the hydrocarbons 
CH$_3$CCH and CH$_2$CCH$_2$ act as precursors of the various nitriles C$_4$H$_3$N found in TMC-1 
\citep{Marcelino2021}, CH$_2$CHCCH emerges as a very likely candidate precursor of the large 
nitriles with molecular formula C$_5$H$_3$N that have recently been claimed in TMC-1 \citep{Lee2021}.

To get insight into the formation mechanism of CH$_2$CHCCH and the other molecules covered 
in this study, we ran a pseudo-time-dependent gas-phase chemical model that adopts typical 
parameters of cold dark clouds (see, e.g., \citealt{Agundez2013}). We used the chemical 
network {\small RATE12} from the UMIST database \citep{McElroy2013}, augmented with reactions 
relevant for the molecules of interest in this work, which are discussed below.

\begin{figure}
\centering
\includegraphics[angle=0,width=0.9\columnwidth]{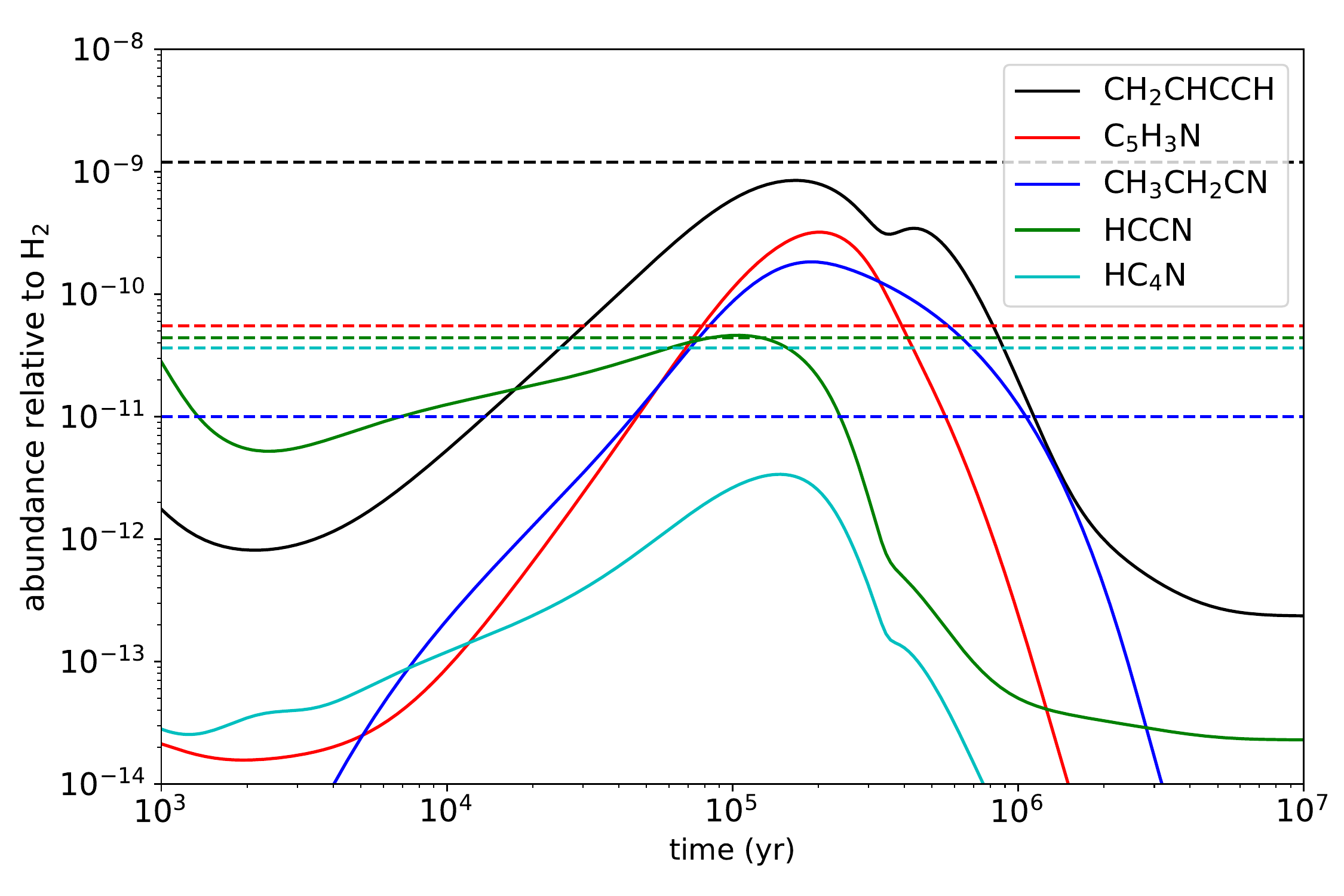}
\caption{Calculated fractional abundances of assorted molecules relevant to this 
study as a function of time. Horizontal dashed lines correspond to observed values.} 
\label{fig:chem_model}
\end{figure}

Vinyl acetylene is formed in the model through the neutral-neutral gas-phase reactions C$_2$ + C$_2$H$_6$, 
CH + C$_3$H$_4$ (where C$_3$H$_4$ stands for the two isomers CH$_3$CCH and 
CH$_2$CCH$_2$), and C$_2$H + C$_2$H$_4$. These reactions have been found to be rapid at low temperatures 
\citep{Canosa2007,Daugey2005,Bowman2012}, and branching ratios have been measured for some of them 
\citep{Loison2009,Goulay2009,Bowman2012}. Vinyl acetylene is predicted to form with a peak abundance
of $\sim 10^{-9}$ relative to 
H$_2$ (see Fig.~\ref{fig:chem_model}), in good agreement with the value derived from 
observations. We also investigated whether cyanide derivatives of CH$_2$CHCCH can 
be formed through the reaction CN + CH$_2$CHCCH, which has been found to be rapid 
for temperatures down to 174 K \citep{Yang1992}. The peak abundance calculated for the 
generic species C$_5$H$_3$N, which accounts for different possible isomers, is a few 
times 10$^{-10}$ relative to H$_2$, somewhat above the observed abundance, which 
means that the reaction between CN and vinyl acetylene is a viable route to 
the C$_5$H$_3$N isomers found in TMC-1 \citep{Lee2021}. Information on the kinetics of this reaction down 
to very low temperatures would be highly valuable to validate this mechanism.

Ethyl cyanide can also be formed efficiently in the gas phase in cold dark cloud conditions. 
In our chemical model the main route involves the radiative association between CH$_3^+$ and 
CH$_3$CN and the dissociative recombination of the ion C$_2$H$_5$CNH$^+$ with electrons, 
both of which have measured rate constants \citep{Anicich1993,Vigren2010}. The calculated 
peak abundance is $\sim 10^{-10}$ relative to H$_2$ (see Fig.~\ref{fig:chem_model}), 
which is around ten times higher than the observed value. 
The chemical network probably misses some important reactions of destruction 
of CH$_3$CH$_2$CN (e.g., with neutral atoms or ions)
which could explain the abundance overestimation.

We finally discuss the formation of the allenic molecules HCCN and 
HC$_4$N in TMC-1. The chemical network that involves HCCN is mostly taken 
from \cite{Loison2015}, while that for HC$_4$N is assumed to be similar. The chemical 
model predicts HCCN to be around ten times more abundant than HC$_4$N 
(see Fig.~\ref{fig:chem_model}), in contrast with observations that find both molecules 
to have similar abundances. The reason for this is that the main routes to HCCN and HC$_4$N are 
the reactions CH + HCN/HNC and CH + HC$_3$N, respectively (with rate constants based 
on \citealt{Zabarnick1991}), and the chemical model predicts that HCN and HNC together 
are around ten times more abundant than HC$_3$N. However, observations indicate that HC$_3$N is as abundant as HCN and HNC in TMC-1 
(e.g., \citealt{Agundez2013}), and therefore if the reactions 
CH + HCN/HNC and CH + HC$_3$N are the main pathways to HCCN and HC$_4$N, respectively, 
one would expect similar abundances for HCCN and HC$_4$N, in agreement with observations. 
The higher HCCN/HC$_4$N observed in IRC\,+10216 is probably explained by the 
higher abundance of HCN compared to HC$_3$N in this source.

\begin{acknowledgements}

We thank Ministerio de Ciencia e Innovaci\'on of Spain (MICIU) for funding support through projects 
AYA2016-75066-C2-1-P, PID2019-106110GB-I00, PID2019-107115GB-C21 / AEI / 10.13039/501100011033, and
PID2019-106235GB-I00. We also thank ERC for funding 
through grant ERC-2013-Syg-610256-NANOCOSMOS. M.A. thanks MICIU for grant RyC-2014-16277.
\end{acknowledgements}

\normalsize

\begin{appendix}
\onecolumn
\section{Line parameters for CH$_2$CHCCH and model fitting procedure}
\label{line_parameters}
Line parameters were derived for CH$_2$CHCCH by fitting a Gaussian line profile
to the observed lines. A velocity range of $\pm$20 \kms around each feature was considered for the fit after a 
polynomial baseline was removed. The derived line parameters are given in Table \ref{tab_line_parameters}. 

For the other species we show the synthetic spectrum resulting from the best fit model to all the lines
computed using the MADEX code \citep{Cernicharo2012}. We assumed a homogeneous rotational temperature
for all rotational levels, a source of uniform brightness with a radius of 40$''$ \citep{Fosse2001}, and
a full linewidth at half power intensity of 0.6 \kms\,, which represents a good averaged value to the
linewidth of all observed lines.
A fit to the observed line profiles and intensities provide the rotational temperature and the column density for the observed
species. The final parameters derived in this way should be very similar to those derived from
a standard rotational diagram. However, this model fit allows us to compare the modeled line
profiles with those of the observations, which is particularly interesting when the rotational transitions
exhibit hyperfine structure, as is the case for HCCN, HC$_4$N, and, to a lesser extent, CH$_3$CH$_2$CN.

\small
\begin{longtable}{lcccccc}
\caption[]{Observed line parameters for CH$_2$CHCCH, HCCN, HC$_4$N, and CH$_3$CH$_2$CN.
\label{tab_line_parameters}}\\
\hline
\hline
Transition      & $\nu$                 & $\int$T$_A^*$dv  & v$_{LSR}$  & $\Delta$v     & T$_A$$^*$  & Notes\\
                &    (MHz)              &(mk km s$^{-1}$)& (km s$^{-1}$)&(km s$^{-1}$)  & (mK)       & \\
\hline
\endfirsthead
\caption{continued.}\\
\hline
Transition      & $\nu$                 &$\int$T$_A^*$dv   & v$_{LSR}$  & $\Delta$v     & T$_A$$^*$  & Notes\\
                &    (MHz)              &(mk km s$^{-1}$)& (km s$^{-1}$)&(km s$^{-1}$)  & (mK)       & \\
\hline
\endhead
\hline
\endfoot
\hline
\endlastfoot
\hline
CH$_2$CHCCH \\
$4_{04}-3_{03}$      & 35463.2066$\pm$0.0003     &  2.1$\pm$0.4& 5.84$\pm$0.04& 0.79$\pm$0.09&  2.5$\pm$0.3 & \\
$4_{14}-3_{13}$      & 36270.1555$\pm$0.0003     &  3.2$\pm$0.5& 5.81$\pm$0.04& 0.87$\pm$0.09&  3.5$\pm$0.4 & \\
$4_{23}-3_{22}$      & 36298.8671$\pm$0.0003     &             &              &              &  .....       & A \\
$4_{22}-3_{21}$      & 36327.0963$\pm$0.0003     &  1.1$\pm$0.4& 5.81$\pm$0.15& 1.22$\pm$0.30&  0.8$\pm$0.3 &  \\
$4_{13}-3_{12}$      & 37123.6151$\pm$0.0003     &  1.0$\pm$0.3& 5.75$\pm$0.04& 0.35$\pm$0.15&  2.7$\pm$0.4 & \\
$5_{15}-4_{14}$      & 44323.5327$\pm$0.0004     &  1.8$\pm$0.3& 5.71$\pm$0.05& 0.59$\pm$0.11&  2.8$\pm$0.5 & \\
$5_{05}-4_{04}$      & 45316.5327$\pm$0.0003     &  1.7$\pm$0.4& 5.87$\pm$0.06& 0.58$\pm$0.15&  2.7$\pm$0.6 & \\
$5_{24}-4_{23}$      & 45369.7261$\pm$0.0003     &             &              &              &  $\le$1.5    & B \\
$5_{23}-4_{22}$      & 45426.1532$\pm$0.0003     &             &              &              &  $\le$1.5    & B \\
$5_{14}-4_{13}$      & 46398.7422$\pm$0.0005     &  2.0$\pm$0.5& 5.76$\pm$0.06& 0.66$\pm$0.15&  2.8$\pm$0.6 & \\
\hline
CH$_3$CH$_2$CN\\
$4_{1,4}-3_{1,3}$ 4-3&  34823.953$\pm$0.001      &  0.9$\pm$0.3& 5.61$\pm$0.13& 0.71$\pm$0.15&  1.2$\pm$0.3 & \\
$4_{1,4}-3_{1,3}$ 5-4&  34824.089$\pm$0.001      &  1.1$\pm$0.3& 5.74$\pm$0.15& 0.80$\pm$0.15&  1.3$\pm$0.3 & \\
$4_{0,4}-3_{0,3}$ 3-2&  35722.143$\pm$0.001      &  1.2$\pm$0.4& 5.83$\pm$0.12& 1.15$\pm$0.30&  1.0$\pm$0.3 & \\
$4_{0,4}-3_{0,3}$ 4-3&  35722.208$\pm$0.001      &  1.8$\pm$0.4& 5.70$\pm$0.06& 0.66$\pm$0.12&  2.5$\pm$0.3 & \\
$4_{0,4}-3_{0,3}$ 5-4&  35722.238$\pm$0.001      &             &              &              &              & NR \\
$4_{1,3}-3_{1,2}$ 5-4&  36739.711$\pm$0.001      &  1.3$\pm$0.4& 5.56$\pm$0.08& 0.83$\pm$0.16&  1.5$\pm$0.3 & BBR\\
$5_{1,5}-4_{1,4}$ 6-5&  43516.229$\pm$0.001      &  1.8$\pm$0.4& 6.00$\pm$0.09& 0.89$\pm$0.20&  1.9$\pm$0.4 & NR\\
$5_{1,5}-4_{1,4}$ 5-4&  43516.153$\pm$0.001      &             &              &              &              & NR\\
$5_{1,5}-4_{1,4}$ 4-3&  43516.186$\pm$0.001      &             &              &              &              & NR\\
$5_{0,5}-4_{0,4}$ 6-5&  44597.010$\pm$0.001      &  2.3$\pm$0.4& 5.79$\pm$0.09& 1.09$\pm$0.18&  1.9$\pm$0.5 & NR\\
$5_{0,5}-4_{0,4}$ 4-3&  44596.954$\pm$0.001      &             &              &              &              & NR\\
$5_{0,5}-4_{0,4}$ 5-4&  44596.986$\pm$0.001      &             &              &              &              & NR\\
$5_{1,4}-4_{1,3}$ 6-5&  45908.544$\pm$0.001      &  1.4$\pm$0.4& 6.09$\pm$0.06& 0.49$\pm$0.14&  2.5$\pm$0.6 & NR\\
$5_{1,4}-4_{1,3}$ 5-4&  45909.472$\pm$0.001      &             &              &              &              & NR\\
$5_{1,4}-4_{1,3}$ 4-3&  45908.514$\pm$0.001      &             &              &              &              & NR\\
\hline
HCCN\\
$1_0-0_1$ 2-1 5/2-5/2&  33681.972$\pm$0.014      &  0.8$\pm$0.3& 5.53$\pm$0.07& 0.49$\pm$0.12&  1.4$\pm$0.3 & \\
$1_1-0_1$ 1-1 3/2-3/2&  33711.125$\pm$0.009      &  0.6$\pm$0.3& 5.74$\pm$0.10& 0.60$\pm$0.15&  0.9$\pm$0.3 & \\
$2_3-1_2$ 3-3 5/2-5/2&  43599.379$\pm$0.004      &  0.9$\pm$0.3& 5.75$\pm$0.12& 0.70$\pm$0.25&  1.2$\pm$0.3 & \\
$2_3-1_2$ 2-2 3/2-3/2&  43609.127$\pm$0.003      &  1.1$\pm$0.4& 5.65$\pm$0.13& 0.96$\pm$0.40&  1.0$\pm$0.4 & \\
$2_3-1_2$ 2-1 3/2-1/2&  43622.442$\pm$0.002      &  1.2$\pm$0.3& 5.60$\pm$0.07& 0.62$\pm$0.12&  1.9$\pm$0.4 & \\
$2_3-1_2$ 2-1 5/2-3/2&  43613.513$\pm$0.002      &             &              &              &              &NR\\
$2_3-1_2$ 4-3 7/2-5/2&  43613.571$\pm$0.002      &  5.0$\pm$0.5& 5.86$\pm$0.02& 0.77$\pm$0.04&  6.2$\pm$0.4 & \\
$2_3-1_2$ 3-2 5/2-3/2&  43624.295$\pm$0.002      &  2.8$\pm$0.4& 5.72$\pm$0.04& 0.70$\pm$0.08&  3.7$\pm$0.4 &\\
$2_3-1_2$ 4-3 9/2-7/2&  43624.662$\pm$0.003      &  2.7$\pm$0.4& 5.74$\pm$0.02& 0.53$\pm$0.06&  4.7$\pm$0.4 &\\
$2_3-1_2$ 3-2 7/2-5/2&  43624.800$\pm$0.002      &  1.6$\pm$0.3& 5.73$\pm$0.03& 0.40$\pm$0.05&  3.8$\pm$0.4 &\\
$2_2-1_1$ 2-1 3/2-1/2&  43936.844$\pm$0.009      &             &              &              &  $\le$1.2    &B\\
$2_2-1_1$ 1-1 1/2-1/2&  43939.743$\pm$0.007      &             &              &              &  $\le$1.2    &B\\
$2_2-1_1$ 3-2 5/2-3/2&  43940.214$\pm$0.006      &  2.2$\pm$0.5& 5.65$\pm$0.09& 1.05$\pm$0.25&  1.9$\pm$0.4 &\\
$2_2-1_1$ 2-2 3/2-3/2&  43943.474$\pm$0.007      &             &              &              &  $\le$1.2    &B\\
$2_2-1_1$ 1-0 3/2-1/2&  43945.275$\pm$0.006      &  1.2$\pm$0.4& 5.45$\pm$0.13& 0.96$\pm$0.30&  1.1$\pm$0.4 &\\
$2_2-1_1$ 2-1 5/2-3/2&  43945.594$\pm$0.003      &  1.0$\pm$0.4& 5.80$\pm$0.11& 0.72$\pm$0.20&  1.3$\pm$0.4 &\\
$2_2-1_1$ 3-2 7/2-5/2&  43948.624$\pm$0.004      &  2.6$\pm$0.4& 5.80$\pm$0.03& 0.66$\pm$0.09&  3.7$\pm$0.4 &\\
\hline
HC$_4$N\\
$ 7_7-6_6$    6- 5 11/2- 9/2& 32233.047$\pm$0.010& 0.5$\pm$0.3& 5.78$\pm$0.20& 0.62$\pm$0.15& 0.8$\pm$0.3&\\
$ 7_7-6_6$    7- 6 13/2-11/2& 32233.149$\pm$0.010& 1.5$\pm$0.4& 5.75$\pm$0.10& 0.98$\pm$0.05& 1.4$\pm$0.3&\\
$ 7_7-6_6$    8- 7 15/2-13/2& 32233.290$\pm$0.003& 1.0$\pm$0.3& 5.80$\pm$0.14& 0.95$\pm$0.09& 1.0$\pm$0.3&\\
$ 7_7-6_6$    6- 5 13/2-11/2& 32233.472$\pm$0.010& 1.6$\pm$0.5& 5.89$\pm$0.08& 1.04$\pm$0.10& 1.4$\pm$0.3&\\
$ 7_7-6_6$    7- 6 15/2-13/2& 32233.545$\pm$0.003&            &              &              &            &NR\\
$ 7_7-6_6$    8- 7 17/2-15/2& 32233.694$\pm$0.003& 0.9$\pm$0.3& 5.83$\pm$0.09& 0.74$\pm$0.18& 1.2$\pm$0.3&\\
$ 7_8-6_7$    7- 6 15/2-13/2& 32365.728$\pm$0.010& 0.9$\pm$0.4& 5.93$\pm$0.11& 0.84$\pm$0.21& 0.9$\pm$0.3&\\
$ 7_8-6_7$    8- 7 17/2-15/2& 32365.938$\pm$0.003&            &              &              &            &NR,BN\\                            
$ 7_8-6_7$    7- 6 13/2-11/2& 32365.972$\pm$0.003& 2.7$\pm$0.5& 5.83$\pm$0.04& 0.87$\pm$0.07& 2.9$\pm$0.3&\\
$ 7_8-6_7$    9- 8 19/2-17/2& 32365.994$\pm$0.003&            &              &              &            &NR,BN\\                            
$ 7_8-6_7$    8- 7 15/2-13/2& 32366.230$\pm$0.010& 1.9$\pm$0.4& 5.51$\pm$0.06& 0.95$\pm$0.13& 1.9$\pm$0.3&\\
$ 7_8-6_7$    9- 8 17/2-15/2& 32366.230$\pm$0.010&            &              &              &            &NR,BN\\
$ 8_7-7_6$    7- 6 13/2-11/2& 36665.590$\pm$0.004& 1.1$\pm$0.3& 5.98$\pm$0.15& 1.00$\pm$0.30& 1.1$\pm$0.3&\\
$ 8_7-7_6$    8- 7 15/2-13/2& 36665.694$\pm$0.004& 0.8$\pm$0.3& 5.74$\pm$0.10& 0.57$\pm$0.17& 1.3$\pm$0.3&\\
$ 8_7-7_6$    6- 5 11/2- 9/2& 36665.831$\pm$0.004& 0.7$\pm$0.3& 5.76$\pm$0.06& 0.32$\pm$0.30& 1.9$\pm$0.3&\\
$ 8_7-7_6$    7- 6 15/2-13/2& 36666.427$\pm$0.004& 2.2$\pm$0.4& 5.64$\pm$0.06& 1.07$\pm$0.14& 2.0$\pm$0.3&\\
$ 8_7-7_6$    8- 7 17/2-15/2& 36666.497$\pm$0.004&            &              &              &            &NR\\
$ 8_7-7_6$    6- 5 13/2-11/2& 36666.641$\pm$0.004& 0.8$\pm$0.4& 5.42$\pm$0.09& 0.68$\pm$0.18& 1.1$\pm$0.3&\\
$ 8_8-7_7$    7- 6 13/2-11/2& 36837.871$\pm$0.005& 2.5$\pm$0.6& 5.65$\pm$0.09& 1.48$\pm$0.25& 1.6$\pm$0.3&\\
$ 8_8-7_7$    8- 7 15/2-13/2& 36837.939$\pm$0.005&            &              &              &            &NR\\
$ 8_8-7_7$    9- 8 17/2-15/2& 36838.043$\pm$0.005&            &              &              &            &NR\\
$ 8_8-7_7$    7- 6 15/2-13/2& 36838.172$\pm$0.005& 1.3$\pm$0.4& 5.66$\pm$0.09& 0.90$\pm$0.22& 1.4$\pm$0.3&\\
$ 8_8-7_7$    8- 7 17/2-15/2& 36838.240$\pm$0.005& 1.3$\pm$0.4& 6.15$\pm$0.05& 0.71$\pm$0.11& 1.7$\pm$0.3&\\ 
$ 8_8-7_7$    9- 8 19/2-17/2& 36838.350$\pm$0.005&            &              &              &            &NR\\
$ 8_9-7_8$    8- 7 17/2-15/2& 36942.973$\pm$0.004&            &              &              & $\le$0.9   &B\\                
$ 8_9-7_8$    8- 7 15/2-13/2& 36943.144$\pm$0.004& 3.2$\pm$0.5& 5.66$\pm$0.03& 0.82$\pm$0.07& 3.7$\pm$0.3&  \\
$ 8_9-7_8$    9- 8 19/2-17/2& 36943.144$\pm$0.004&            &              &              &            &NR\\
$ 8_9-7_8$   10- 9 21/2-19/2& 36943.179$\pm$0.004&            &              &              &            &NR\\
$ 8_9-7_8$    9- 8 17/2-15/2& 36943.330$\pm$0.004& 1.5$\pm$0.4& 5.67$\pm$0.04& 0.60$\pm$0.08& 2.4$\pm$0.3&\\
$ 8_9-7_8$   10- 9 19/2-17/2& 36943.352$\pm$0.004&            &              &              &            &NR\\
$ 9_8-8_7$    8- 7 15/2-13/2& 41309.798$\pm$0.007& 1.3$\pm$0.3& 5.94$\pm$0.15& 1.19$\pm$0.12& 1.0$\pm$0.3&\\
$ 9_8-8_7$    9- 8 17/2-15/2& 41309.886$\pm$0.007&            &              &              &            &NR\\
$ 9_8-8_7$    7- 6 13/2-11/2& 41309.971$\pm$0.007& 2.4$\pm$0.5& 5.67$\pm$0.10& 0.33$\pm$0.04& 0.7$\pm$0.3&\\
$ 9_8-8_7$    8- 7 17/2-15/2& 41310.387$\pm$0.007&            &              &              &            &NR\\
$ 9_8-8_7$    9- 8 19/2-17/2& 41310.452$\pm$0.007& 1.3$\pm$0.4& 5.85$\pm$0.06& 0.64$\pm$0.09& 1.9$\pm$0.3&\\
$ 9_8-8_7$    7- 6 15/2-13/2& 41310.540$\pm$0.007&            &              &              &            &NR\\
$ 9_9-8_8$    8- 7 15/2-13/2& 41442.624$\pm$0.007&            &              &              &            &NR\\
$ 9_9-8_8$    9- 8 17/2-15/2& 41442.703$\pm$0.007&            &              &              &            &NR\\
$ 9_9-8_8$   10- 9 19/2-17/2& 41442.760$\pm$0.007&            &              &              &            &NR\\
$ 9_9-8_8$    8- 7 17/2-15/2& 41442.839$\pm$0.007& 2.4$\pm$0.6& 5.54$\pm$0.30& 2.44$\pm$0.50& 0.9$\pm$0.3&\\
$ 9_9-8_8$    9- 8 19/2-17/2& 41442.917$\pm$0.007&            &              &              &            &NR\\
$ 9_9-8_8$   10- 9 21/2-19/2& 41443.002$\pm$0.007&            &              &              &            &NR\\
$ 9_{10}-8_9$   9- 8 19/2-17/2& 41527.729$\pm$0.007& 1.0$\pm$0.3& 5.68$\pm$0.06& 0.51$\pm$0.12& 1.8$\pm$0.4&\\
$ 9_{10}-8_9$   9- 8 17/2-15/2& 41527.847$\pm$0.007&            &              &              &            &NR\\
$ 9_{10}-8_9$  10- 9 21/2-19/2& 41527.865$\pm$0.007&            &              &              &            &NR\\
$ 9_{10}-8_9$  11-10 23/2-21/2& 41527.887$\pm$0.007& 2.7$\pm$0.6& 5.90$\pm$0.04& 0.70$\pm$0.11& 3.6$\pm$0.4&\\
$ 9_{10}-8_9$  10- 9 19/2-17/2& 41527.994$\pm$0.007& 0.6$\pm$0.3& 6.86$\pm$0.12& 0.53$\pm$0.29& 1.0$\pm$0.4&\\
$ 9_{10}-8_9$  11-10 21/2-19/2& 41528.006$\pm$0.007&            &              &              &            &NR\\
$10_{10}-9_9$   9- 8 17/2-15/2& 46047.335$\pm$0.010&            &              &              &            &NR\\
$10_{10}-9_9$   9- 9 19/2-17/2& 46047.440$\pm$0.010&            &              &              &            &NR\\
$10_{10}-9_9$  11-10 21/2-19/2& 46047.445$\pm$0.010& 2.5$\pm$0.4& 5.86$\pm$0.08& 0.84$\pm$0.17& 2.8$\pm$0.6&\\
$10_{10}-9_9$  10- 8 19/2-17/2& 46047.469$\pm$0.010&            &              &              &            &NR\\
$10_{10}-9_9$  10- 9 21/2-19/2& 46047.573$\pm$0.010&            &              &              &            &NR\\
$10_{10}-9_9$  11-10 23/2-21/2& 46047.641$\pm$0.010& 2.0$\pm$0.4& 6.01$\pm$0.06& 0.57$\pm$0.13& 3.2$\pm$0.6&\\
$10_{11}-9_{10}$ 10- 9 21/2-19/2& 46117.484$\pm$0.010&            &              &              &            &NR\\
$10_{11}-9_{10}$ 10- 9 19/2-17/2& 46117.567$\pm$0.010&            &              &              &            &NR\\
$10_{11}-9_{10}$ 11-10 23/2-21/2& 46117.594$\pm$0.010& 1.6$\pm$0.5& 5.74$\pm$0.06& 0.51$\pm$0.12& 3.3$\pm$0.6&\\
$10_{11}-9_{10}$ 12-11 25/2-23/2& 46117.608$\pm$0.010&            &              &              &            &NR\\
$10_{11}-9_{10}$ 11-10 21/2-19/2& 46117.685$\pm$0.010& 1.3$\pm$0.5& 5.70$\pm$0.11& 0.62$\pm$0.17& 2.0$\pm$0.6&\\
$10_{11}-9_{10}$ 12-11 23/2-21/2& 46117.692$\pm$0.010&            &              &              &            &NR\\
\hline
CH$_2$CHCN\\
$4_{1,4}-3_{1,3}$ 4-4&  37017.816$\pm$0.001&  1.7$\pm$0.3& 5.87$\pm$0.05& 0.68$\pm$0.11&  2.4$\pm$0.4 &\\
$4_{1,4}-3_{1,3}$ 4-3&  37018.829$\pm$0.001& 26.8$\pm$0.8& 5.77$\pm$0.01& 0.69$\pm$0.02& 36.3$\pm$0.4 &\\
$4_{1,4}-3_{1,3}$ 3-2&  37018.926$\pm$0.001& 22.8$\pm$0.7& 5.73$\pm$0.02& 0.68$\pm$0.05& 31.4$\pm$0.4 &\\
$4_{1,4}-3_{1,3}$ 5-4&  37018.981$\pm$0.001& 29.3$\pm$0.8& 5.74$\pm$0.01& 0.64$\pm$0.02& 42.9$\pm$0.4 &\\
$4_{1,4}-3_{1,3}$ 3-3&  37020.293$\pm$0.001&  0.9$\pm$0.3& 5.72$\pm$0.15& 0.51$\pm$0.34&  1.6$\pm$0.4 &\\
$4_{0,4}-3_{0,3}$ 4-4&  37903.587$\pm$0.001&  3.0$\pm$0.4& 5.81$\pm$0.03& 0.66$\pm$0.08&  4.3$\pm$0.4 &\\
$4_{0,4}-3_{0,3}$ 3-2&  37904.770$\pm$0.001& 31.7$\pm$0.8& 5.87$\pm$0.01& 0.69$\pm$0.02& 42.8$\pm$0.4 &\\
$4_{0,4}-3_{0,3}$ 4-3&  37904.850$\pm$0.001& 87.2$\pm$1.5& 5.75$\pm$0.02& 0.72$\pm$0.01&113.8$\pm$0.4 &\\
$4_{0,4}-3_{0,3}$ 5-4&  37904.880$\pm$0.001&             &              &              &              &NR\\
$4_{0,4}-3_{0,3}$ 3-3&  37906.474$\pm$0.001&  1.5$\pm$0.4& 5.72$\pm$0.03& 0.49$\pm$0.12&  2.9$\pm$0.4 &\\
$4_{2,3}-3_{2,2}$ 4-3&  37939.248$\pm$0.001&  4.2$\pm$0.5& 5.90$\pm$0.01& 0.62$\pm$0.03&  6.3$\pm$0.4 &\\
$4_{2,3}-3_{2,2}$ 5-4&  37939.763$\pm$0.001&  6.2$\pm$0.5& 5.78$\pm$0.01& 0.74$\pm$0.04&  7.8$\pm$0.4 &\\
$4_{2,3}-3_{2,2}$ 3-2&  37939.896$\pm$0.001&  2.6$\pm$0.4& 5.79$\pm$0.02& 0.59$\pm$0.05&  4.1$\pm$0.4 &\\
$4_{2,2}-3_{2,1}$ 4-3&  37973.989$\pm$0.001&  4.7$\pm$0.4& 5.83$\pm$0.02& 0.74$\pm$0.04&  6.0$\pm$0.4 &\\
$4_{2,2}-3_{2,1}$ 5-4&  37974.504$\pm$0.001&  5.5$\pm$0.5& 5.80$\pm$0.01& 0.68$\pm$0.03&  7.6$\pm$0.4 &\\
$4_{2,2}-3_{2,1}$ 3-2&  37974.636$\pm$0.001&  3.1$\pm$0.4& 5.82$\pm$0.02& 0.65$\pm$0.05&  4.4$\pm$0.4 &\\
$4_{1,3}-3_{1,2}$ 4-4&  38846.762$\pm$0.001&  1.5$\pm$0.4& 5.88$\pm$0.06& 0.89$\pm$0.13&  1.6$\pm$0.4 &\\
$4_{1,3}-3_{1,2}$ 4-3&  38847.641$\pm$0.001& 27.4$\pm$0.8& 5.75$\pm$0.00& 0.66$\pm$0.01& 38.9$\pm$0.4 &\\
$4_{1,3}-3_{1,2}$ 3-2&  38847.747$\pm$0.001& 55.4$\pm$1.1& 5.88$\pm$0.00& 0.79$\pm$0.01& 66.3$\pm$0.4 &\\
$4_{1,3}-3_{1,2}$ 5-4&  38847.790$\pm$0.001&             &              &              &              &NR\\ 
$4_{1,3}-3_{1,2}$ 3-3&  38848.933$\pm$0.001&  2.5$\pm$0.4& 5.81$\pm$0.02& 0.63$\pm$0.06&  3.4$\pm$0.6 &\\ 
$5_{1,5}-4_{1,4}$ 5-4&  46266.887$\pm$0.001&             &              &              &              &NR\\
$5_{1,5}-4_{1,4}$ 4-3&  46266.926$\pm$0.001& 70.2$\pm$1.2& 5.76$\pm$0.01& 0.87$\pm$0.01& 76.0$\pm$0.6 &\\
$5_{1,5}-4_{1,4}$ 6-5&  46266.971$\pm$0.001&             &              &              &              &NR\\
$5_{0,5}-4_{0,4}$ 5-5&  47353.356$\pm$0.001&  1.9$\pm$0.4& 5.80$\pm$0.05& 0.56$\pm$0.11&  3.2$\pm$0.6 &\\
$5_{0,5}-4_{0,4}$ 4-3&  47354.605$\pm$0.001&             &              &              &              &NR\\
$5_{0,5}-4_{0,4}$ 5-4&  47354.648$\pm$0.001&104.0$\pm$1.5& 5.80$\pm$0.01& 0.74$\pm$0.01&132.9$\pm$0.6 &\\
$5_{0,5}-4_{0,4}$ 6-5&  47354.670$\pm$0.001&             &              &              &              &NR\\
$5_{0,5}-4_{0,4}$ 4-4&  47356.230$\pm$0.001&  2.5$\pm$0.4& 6.00$\pm$0.06& 0.75$\pm$0.09&  3.1$\pm$0.6 &\\
$5_{2,4}-4_{2,3}$ 5-4&  47419.606$\pm$0.001&  5.5$\pm$0.6& 5.77$\pm$0.02& 0.69$\pm$0.05&  7.6$\pm$0.6 &\\
$5_{2,4}-4_{2,3}$ 6-5&  47419.876$\pm$0.001&  8.6$\pm$0.8& 5.75$\pm$0.02& 0.70$\pm$0.05& 11.5$\pm$0.6 &\\
$5_{2,4}-4_{2,3}$ 4-3&  47419.904$\pm$0.001&             &              &              &              &NR\\
$5_{1,4}-4_{1,3}$ 5-4&  48552.516$\pm$0.001&             &              &              &              &NR\\
$5_{1,4}-4_{1,3}$ 4-3&  48552.559$\pm$0.001& 73.8$\pm$1.9& 5.77$\pm$0.01& 0.86$\pm$0.02& 80.3$\pm$0.9 &\\
$5_{1,4}-4_{1,3}$ 6-5&  48552.598$\pm$0.001&             &              &              &              &NR\\
\hline
\end{longtable}
\tablefoot{\\
\tablefoottext{A}{Blended with a strong feature.}\\
\tablefoottext{B}{Three sigma upper limit.}\\
\tablefoottext{NR}{Hyperfine structure line blended with other components of
the same rotational transition. The line has been included in the fit to the previous
or next entry in the table.}\\
}
\normalsize

\section{Ethyl acetylene and ethyl cyanide}
\label{CH3CH2CCH}
\begin{figure*}[]
\centering
\includegraphics[scale=0.85,angle=0]{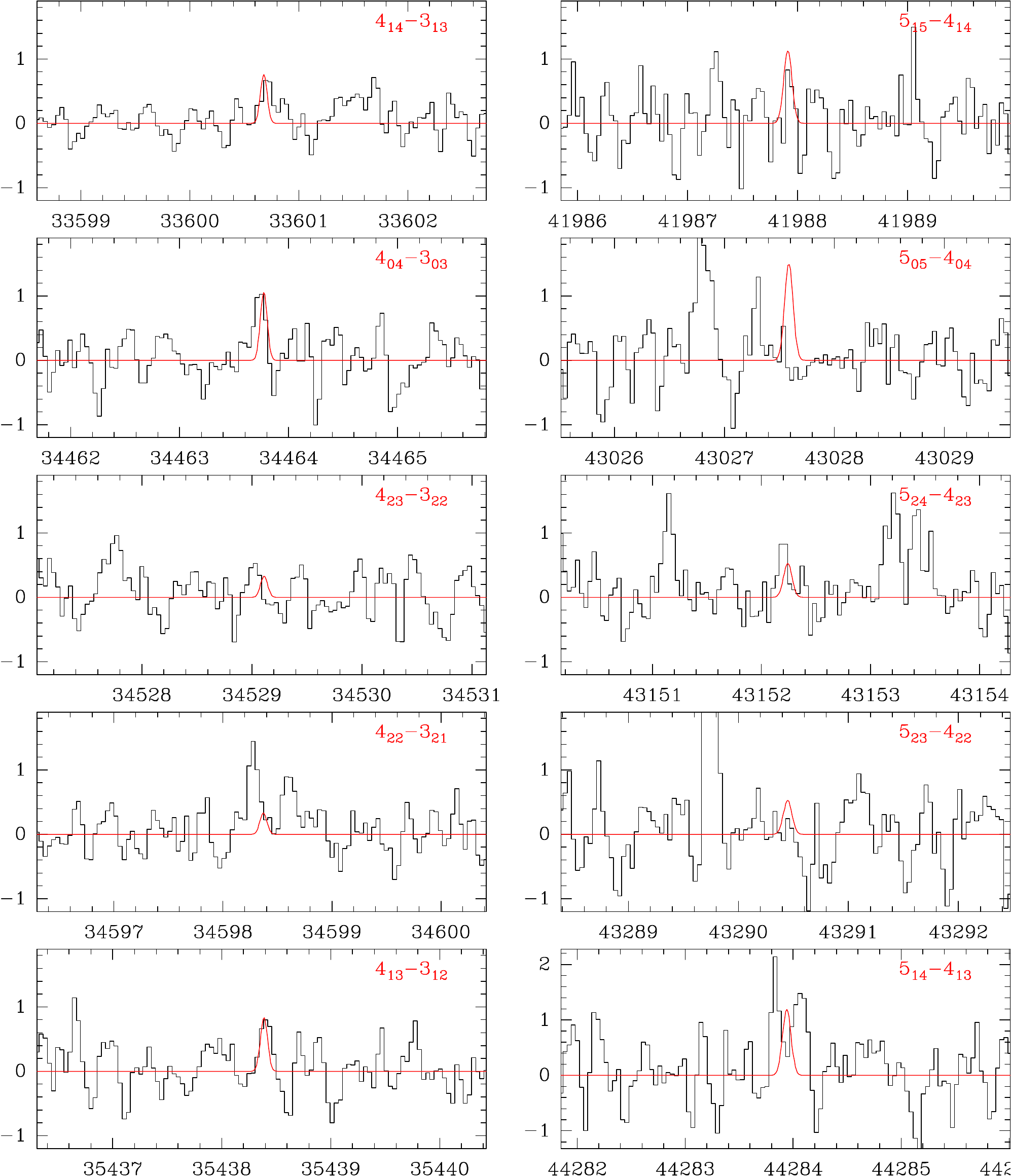}
\caption{Observed lines of CH$_3$CH$_2$CCH in the Q-band toward TMC-1.
The abscissa corresponds to the rest frequency assuming a local standard of rest velocity of 5.83
km s$^{-1}$. 
The ordinate is the antenna temperature, corrected for atmospheric and telescope losses, in mK.
The red line shows the synthetic spectrum obtained using the parameters derived from a
rotational diagram of the observed lines (T$_r$=5 K and 
N(CH$_3$CH$_2$CCH)=9.0$\times$10$^{11}$ cm$^{-2}$).
The rotational quantum
numbers are provided in the upper right corners of each panel. 
}
\label{fig_ch3ch2cch}
\end{figure*}

We searched for the lines of ethyl acetylene, CH$_3$CH$_2$CCH, a molecule for which accurate laboratory data 
are available up
to 317.7 GHz \citep{Demaison1983,Landsberg1983,Bestmann1985,Steber2012} and which has moderate dipole moments 
of $\mu_a$=0.763 D and $\mu_b$=0.17 D \citep{Landsberg1983}. 
The barrier to internal rotation is very high \citep{Bestmann1985}, and therefore 
the ground state does not show significant splitting between the $A$ and $E$ species.
Of the ten lines expected for this molecule in the Q-band, five are detected at a 3$\sigma$ level, two are blended with other
weak features slightly shifted in frequency, and three are too weak.
Using stacking techniques (see below) and removing the blended lines, which could introduce a significant bias in
the final spectrum, we detected a signal at 5$\sigma$ at the correct
velocity, as shown in Fig. \ref{fig_ch3ch2cch_sta}. However, we have no
explanation for the lack of emission at the frequency of the $5_{05}-4_{04}$ transition other than that the data are too noisy at
its frequency (see Fig. \ref{fig_ch3ch2cch}). It is expected to be the strongest feature
for the parameters we used for the synthetic spectrum (T$_r$=5 K, N(CH$_3$CH$_2$CCH)=9 10$^{11}$ \cd).

For CH$_3$CH$_2$CCH, all the expected strongest lines in the Q-band are shown in Fig. \ref{fig_ch3ch2cch}, 
and Fig. \ref{fig_ch3ch2cch_sta} shows the resulting
stacked profile. The stacking procedure we used is rather simple: A velocity range of $\pm$20 \kms\  was selected
for each line and the noise, $\sigma$, outside 5.83$\pm$0.5 \kms\, was computed (with known and unknown lines
in each individual spectrum blanked). The data were normalized to the 
expected intensity of each transition computed
under local thermodynamic equilibrium for a rotational temperature of 5 K, the same as the value
derived for CH$_2$CHCCH. Finally, all the data were multiplied by the expected intensity of
the strongest transition in the sample. The lines are optically thin, and hence the intensity of all lines scale in the same way
with the assumed column density. Each individual spectrum was weighted as 1/$\sigma_N^2$, where $\sigma_N$ now contains the normalization intensity factor. 
All rotational
transitions falling in the frequency ranges where our data have the highest sensitivity were detected (transitions
$4_{14}-3_{13}, 4_{04}-3_{03}, 4_{13}-3_{12}, 5_{15}-4_{14}$, and $5_{24}-5_{23}$). 
Two of the observed transitions are
heavily blended ($4_{22}-3_{21}$ and $5_{14}-4_{13}$) and were excluded from the stacked spectrum. 
Two $K_a$=2 transitions, which are expected to be weak, were not detected ($4_{23}-3_{22}$ and $5_{23}-4_{22}$) but were included
in the data stacking with the corresponding weights, as explained above.
The data for the strongest rotational transition, the $5_{05}-4_{04}$, have a sensitivity of 0.6 mK. The expected
intensity is 1.3 mK. This rotational transition was included in the stacked spectrum, although it is clearly not detected at the noise
level of the data. Figure \ref{fig_ch3ch2cch_sta} shows the resulting spectrum from this procedure. A feature, at exactly the
velocity of the cloud, appears at a 5$\sigma$ level ($\sigma$=0.16 mK). 
Although a positive detection of ethyl acetylene is highly possible, we consider it as tentative.
Longer integration times are needed to confirm the detection of this species. 

\begin{figure}[]
\centering
\includegraphics[scale=0.6,angle=0]{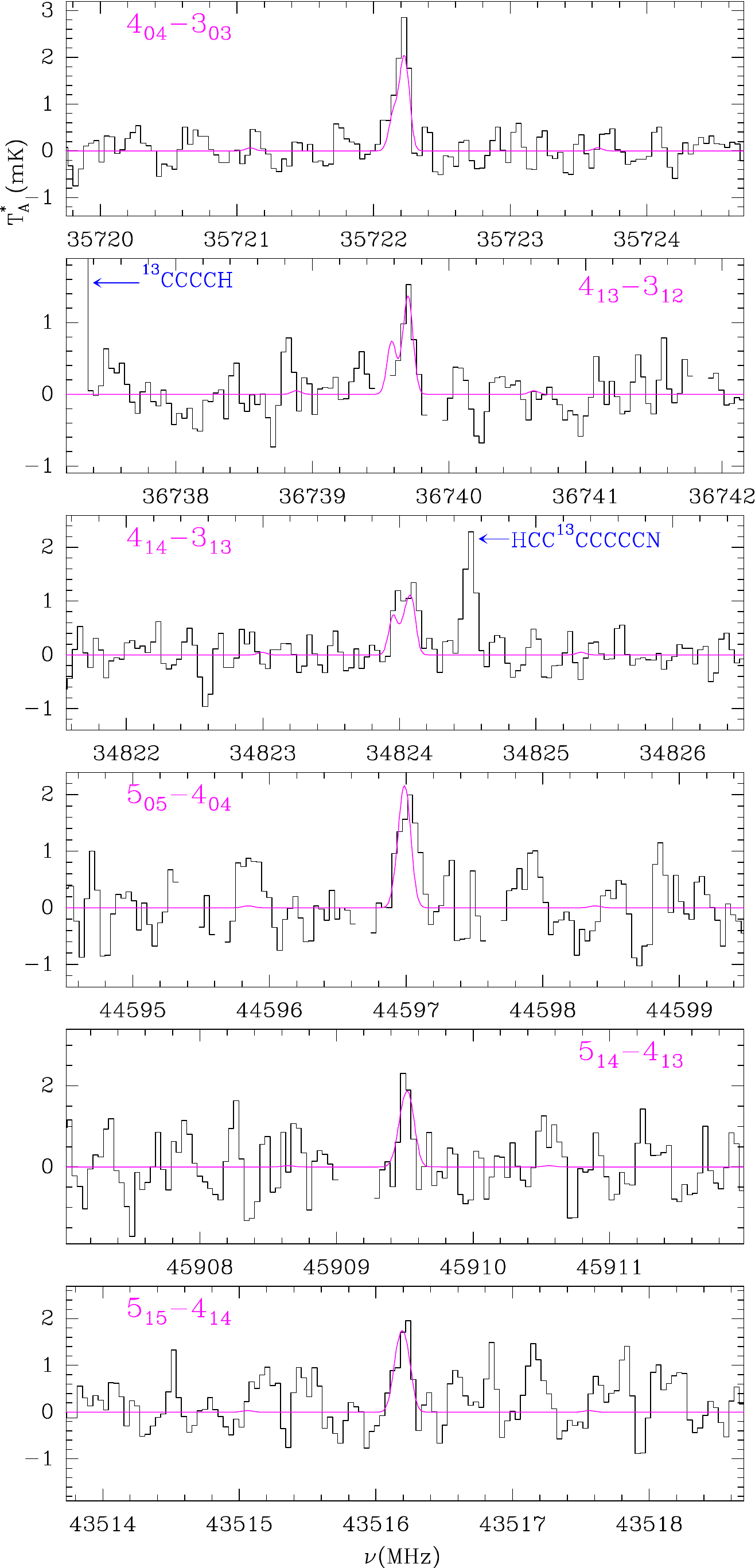}
\caption{Observed lines of CH$_3$CH$_2$CN in the 31-50 GHz frequency range toward TMC-1.
The abscissa corresponds to the rest frequency assuming a local standard of rest velocity of 5.83
km s$^{-1}$. The ordinate is the antenna temperature, corrected for atmospheric and telescope losses, in mK.
The violet line shows the synthetic spectrum computed for T$_r$=4.5 K and N(CH$_3$CH$_2$CN)=1.1$\times$\once.}
\label{fig_ch3ch2cn}
\end{figure}

In our survey we observed a large number of unidentified lines at the level of 1-5 mK (see Figs. 1, 3, 4, and 
\ref{fig_ch3ch2cch}). 
Moreover, the isotopologues of abundant species contribute with several hundred lines. Consequently,
stacking techniques that could compete in sensitivity with our survey may provide false positive or negative 
results for low abundant molecular species. An example is the case of CH$_3$CH$_2$CN, which was detected
in our work (see Sect. \ref{CH3CH2CN} and Fig. \ref{fig_ch3ch2cn}), while only upper limits were obtained by \citet{Lee2021}.
The observed lines of CH$_3$CH$_2$CN, together with the
best model fit to the observed emission, are shown in Fig. \ref{fig_ch3ch2cn}.

\section{Vinyl cyanide, CH$_2$CHCN}
\begin{figure*}[]
\centering
\includegraphics[scale=0.95,angle=0]{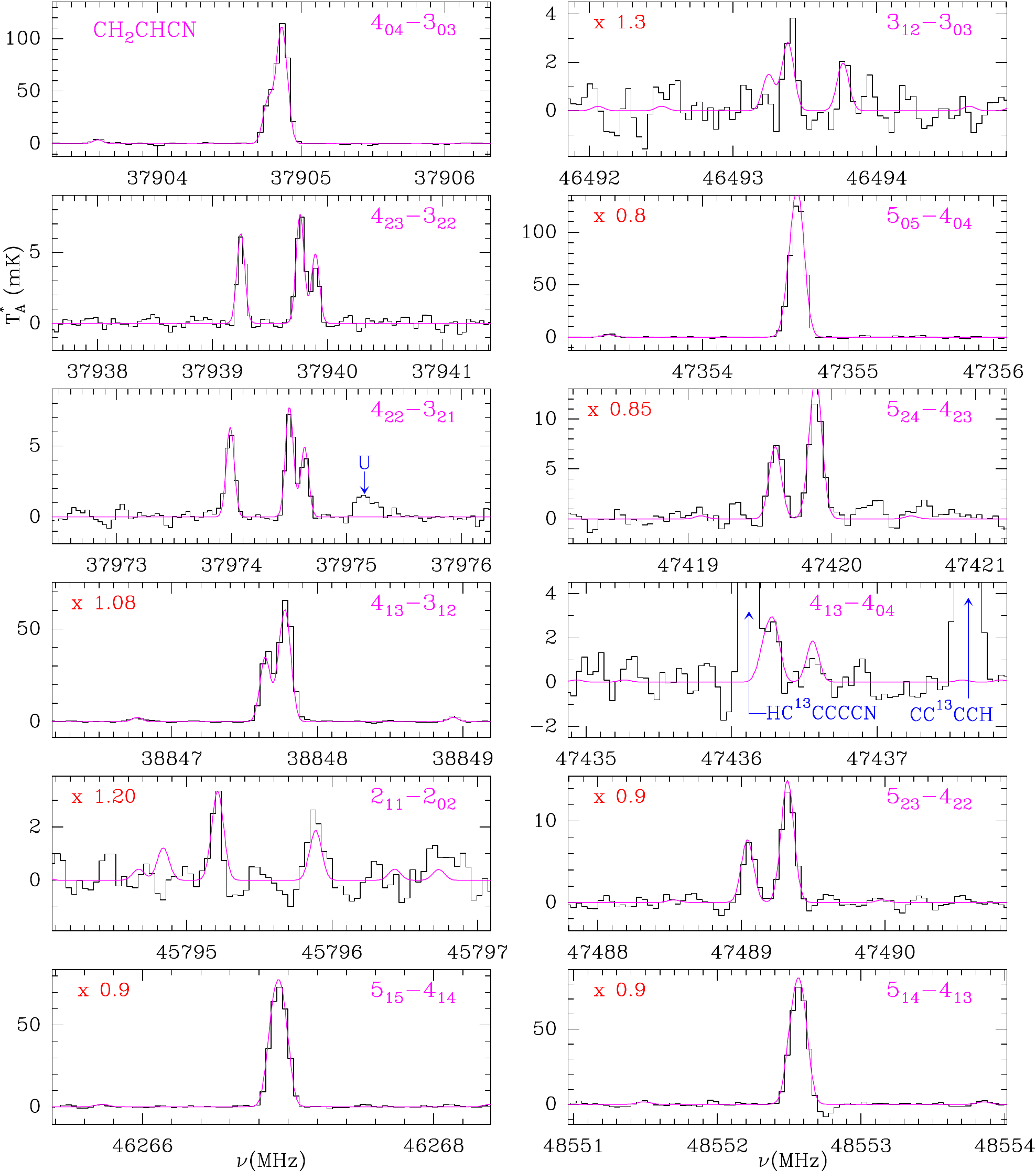}
\caption{Observed lines of CH$_2$CHCN in the 31-50 GHz frequency range toward TMC-1.
The abscissa corresponds to the rest frequency assuming a local standard of rest velocity of 5.83
km s$^{-1}$. 
The ordinate is the antenna temperature, corrected for atmospheric and telescope losses, in mK.
The violet line shows the synthetic spectrum obtained using the parameters derived from a rotational 
diagram of the observed lines (T$_r$=4.5$\pm$0.3 K and 
N(CH$_2$CHCN)=(6.5$\pm$0.5)$\times$10$^{12}$ cm$^{-2}$). 
Numbers in the upper left corners of the 
panels indicate the multiplicative factor applied to the synthetic spectrum to match the data.
The rotational quantum
numbers are provided in the upper right corners of each panel. The hyperfine structure of each rotational
transition has been taken into account in the calculation of the synthetic spectra. Line parameters
are given in Table \ref{tab_line_parameters}.
}
\label{fig_ch2chcn}
\end{figure*}

Vinyl cyanide produces a very rich spectrum in the 31-50 GHz range, with most of its lines
exhibiting the hyperfine structure due to the nitrogen nucleus. All  the observed lines
are shown in Fig. \ref{fig_ch2chcn}. The model fitting procedure provides a rotational
temperature of 4.5$\pm$0.5 K and a column density of (6.5$\pm$0.5)$\times$10$^{12}$ cm$^{-2}$.
Its abundance relative to vinyl acetylene and 
ethyl cyanide is discussed in Sects. \ref{CH2CHCCH}, \ref{CH3CH2CN}, and \ref{discussion}.

\label{CH2CHCN}

\end{appendix}

\end{document}